\documentclass[aps,nofootinbib,preprintnumbers,showpacs,prd,twocolumn,superscriptaddress]{revtex4-1}
\usepackage{array}
\usepackage{booktabs}
\usepackage{subfigure}
\usepackage{capt-of}
\usepackage{graphicx,amsmath,bm}
\usepackage{lipsum}
\usepackage{epstopdf}
\usepackage{amsmath}
\usepackage{amssymb}
\usepackage{color,xcolor}
\usepackage[bookmarks=false]{hyperref}
\hypersetup{colorlinks=true,citecolor=green,linkcolor=blue,urlcolor=blue,pdfstartview=FitH,bookmarksopen=true}

\begin{document}
	
\title{Sivers function of sea quarks in the Light-Cone Model}
\author{Xiaoyan Luan}
\affiliation{School of Physics, Southeast University, Nanjing 211189, China}
\author{Zhun Lu}\email{zhulu@seu.edu.cn}
\affiliation{School of Physics, Southeast University, Nanjing 211189, China}

\begin{abstract}
We calculate the Sivers function of $\bar{u}$ and $\bar{d}$ quarks using the overlap representation within the light-cone formalism.
The light-cone wave functions of the proton is obtained in terms of the $|\bar{q} q B\rangle$ Fock states motivated by the meson-baryon fluctuation model.
We consider the final-state interaction at the level of one gluon exchange.
In a simplified scenario, the  Sivers function of  $\bar{u}$ and $\bar{d}$  can be expressed as the convolution of the Sivers function of the pion inside the proton and the unpolarized distribution of $\bar{q}$ inside the pion.
The model parameters are fixed by fitting the unpolarized sea quark distributions to the known parameterizations.
We present the numerical results for $f_1^{\perp \bar{u}/p}(x,\boldsymbol{k}_T^2)$ and $f_1^{\perp \bar{d}/p}(x,\boldsymbol{k}_T^2)$.
The first transverse moment of the sea quark Sivers functions in our model are find to be negative and the magnitude is about 0.004 at most.
\end{abstract}

 \maketitle

\section{Introduction}\label{Sec1}

The transverse momentum dependent (TMD) parton distribution functions (PDFs), also called as unintegrated PDFs, describe the probability of finding a quark with certain light-cone momentum fraction $x$ and transverse momentum $k_T$ in a hadron~\cite{collins1982parton}.
Among them the Sivers function~\cite{Sivers:1989cc} is of particular interests.
Due to the time-reversal-odd (T-odd) spin-orbit correlation of partons inside the nucleon,
the Sivers function describes the distortion in the distribution of unpolarized partons inside a transversely polarized nucleon.
Thereby it can give rise to the single-spin asymmetries (SSAs) in leading-twist in various semi-inclusive high energy processes.
During the last two decades, the SSAs related to Sivers function in semi-inclusive deep inelastic scattering (SIDIS) have been measured by the HERMES~\cite{Airapetian:2004tw,Airapetian:2009ae}, COMPASS~\cite{COMPASS:2008isr,COMPASS:2010hbb,COMPASS:2012dmt,COMPASS:2016led}, and JLab Hall A~\cite{Qian:2011py,JeffersonLabHallA:2014yxb} Collaborations.
Recently, the measurement on the Sivers effects in Drell-Yan and W/Z production have also been performed by COMPASS~\cite{COMPASS:2017jbv} and STAR~\cite{STAR:2015vmv}.
The data from these experiments were utilized by
different groups~\cite{Anselmino:2005ea,Efremov:2004tp,Collins:2005ie,Vogelsang:2005cs,Anselmino:2008sga,
Anselmino:2012aa,bacchetta2011constraining,Echevarria:2014xaa,anselmino2017study, martin2017direct, boglione2018assessing,Bury:2021sue}
to extract the quark Sivers functions of the proton without or with TMD evolution.
On the other hand, various model calculations have been applied to calculate the quark Sivers functions, such as the the spectator model~\cite{brodsky2002final, boer2003initial, bacchetta2004sivers}, the light-cone quark model~\cite{lu2004nonzero, pasquini2010sivers}, the non-relativistic constituent quark model~\cite{Courtoy:2008vi}, and the MIT bag model~\cite{yuan2003sivers,Courtoy:2008dn}.

Although there are plenty of measurements on the Sivers asymmetry combined with phenomenological extractions, our knowledge on the Sivers function is still limited. 
Currently, most of the information from theoretical study and data focuses on the Sivers functions of the valence quarks.
The Sivers functions of the sea quarks and gluon are less constrained compared to those of the $u$ and $d$ quarks.
In the Drell-Yan process, the anti-quarks play a more significant role since the dominant sub-process is $q \bar{q}\rightarrow \gamma^\star\rightarrow$.
The time-reversal property of the Sivers function predicts that it present generalized universality, i.e., the sign of the Sivers function measured in Drell-Yan process should be opposite to its sign measured in SIDIS~\cite{Collins:2002kn,Brodsky:2002cx,Brodsky:2002rv}.
The verification of this sign change~\cite{Anselmino:2009st,Kang:2009bp,Peng:2014hta,Echevarria:2014xaa,Anselmino:2016uie} is one of the most fundamental tests of our understanding of the QCD dynamics, and it is also the main pursue of the existing and future Drell-Yan facilities~\cite{COMPASS:2017jbv,STAR:2015vmv,Arbuzov:2020cqg}.
As pointed out in Ref.~\cite{Bury:2021sue}, the sea quark Sivers function could play an important role in the test of the sign-change between SIDIS and Drell-Yan, since the sign for anti-quark Sivers function almost exclusively defines the sign of the asymmetry of Drell-Yan and $W/Z$ production in polarized $p p$ collision.
Furthermore, sea quark Sivers function is also of importance to explain the azimuthal asymmetry of kaon production in in SIDIS.

In this work, we apply an intuitive model to calculate the Sivers functions of the $\bar{u}$ and $\bar{d}$ quarks using the overlap representation.
In the model the sea quark degree freedom is generated by the assumption that the proton can fluctuate to a composite state containing a meson $M$ and a baryon $B$.
It is similar to the meson cloud effect which was proposed by Sullivan through analyzing the deep inelastic scattering (DIS) process~\cite{sullivan1972one}, 
and has been applied to study the sea quark distributions of the nucleon~\cite{koepf1996virtual,huang2004sigma,brodsky1996quark}.
We consider the $\bar{q}q$ component of the pion meson and derive the light-cone wave functions (LCWFs) of the proton in terms of the Fock-state $|\bar{q} q B\rangle$.
As proposed in Ref.~\cite{Lu:2006kt}, the overlap representation~\cite{Brodsky:2000ii} can be also applied to calculate the Sivers function with LCWFs.
In this approach, the Sivers function is proportional to the overlap of the LCWFs of the initial-state and final-state nucleons differ by $\Delta L_z=\pm 1$, where $L_z$ is the orbital angular momentum of the active quark/antiquark.
The values of the parameters in the model are determined by fitting the unpolarized sea quark distributions in the same model to the available parameterizations.
We note that the Sivers function of the light sea has been calculated with a chiral Lagrangian~\cite{he2019sivers};
the sea quark Sivers functions in the small $x$ region has also been studied within color glass condensate framework~\cite{Dong:2018wsp}.

The paper is organized as follows.
In Sec.~II, we derive the LCWFs of the proton in terms of the Fock state $|\bar{q}q B\rangle$.
In Sec.~III, we apply the LCWFs to calculate the unpolarized sea quark distribution $f_1^{\bar{q}/p}(x,\bm k_T^2)$ and the sea quark Sivers function $f_{1T}^{\bot\bar{q}/p}(x,\bm k_T^2)$.
In Sec.~IV, we present the numerical result for the sea quark Sivers function $f_{1T}^{\bot\bar{q}/p}(x,\bm k_T^2)$ and the first transverse momentum $f_{1T}^{\bot(1)\bar{q}/p}(x)$.
We summarize the paper in Sec.~V.

\section{LCWFs of the proton in terms of $|\bar{q}q B\rangle$.}
\label{Sec2}

The light-cone formalism has been widely recognized as a convenient way to calculate the parton distribution functions of nucleon and meson~\cite{Lepage:1980aa}.
Within the light-cone approach, the wave functions for a hadronic composite state can be expressed as LCWFs in Fock-state basis.
On the other hand, the overlap representation has also been used to study various form factors of the hadrons~\cite{Brodsky:2000ii} and the pion~\cite{Xiao:2003wf}, anomalous magnetic moment of the nucleon~\cite{Brodsky:2000ii} as well as generalized parton distribution functions~\cite{Brodsky:2000xy}. recently, the has also applied to calculate the T-odd TMDs ~\cite{Lu:2006kt,Bacchetta:2008af}
In this section, we will extend the light-cone formalism to calculate the Sivers function of sea quarks.

Another important ingredient needed in the calculation is the generation of the sea quark degree of freedom. 
For this purpose we will apply the baryon-meson fluctuation model, in which the the proton can fluctuate to a composite system formed by a meson $M$ and a baryon $B$, where the meson is composed in terms of $q\bar{q}$:
\begin{align}\label{eq1}
|p\rangle\to| M B\rangle\to|q\bar{q}B\rangle.
\end{align}
In this work we consider the fluctuation $|p\rangle \to |\pi^+ n\rangle $ and
$|p\rangle \to |\pi^- \Delta^{++}\rangle $.
For the above proton composite state, the LCWFs can be calculated from the following matrix elements~\cite{Brodsky:2000ii,Xiao:2003wf,Bacchetta:2008af}
\begin{align}\label{eq2}
\notag&
\psi^{\lambda_N}_{{\lambda_B}{\lambda_q}{\lambda_{\bar{q}}}}(r,k)
\\\notag=&\sqrt{\frac{r^+}{(P-r)^+}}\frac{1}{r^2-m_\pi^2}\overline{U}_B(P-r,\lambda_B)\gamma_5U(P,\lambda_P)
\nonumber\\
&\notag\times\sqrt{\frac{k^+}{(r-k)^+}}\frac{1}{k^2-m^2}\overline{u}(r-k,\lambda_q)
\gamma_5\nu(k,\lambda_{\bar{q}})\\
=&\psi^{\lambda_N}_{\lambda_B}(y,\boldsymbol{r}_T)\psi_{{\lambda_q}{\lambda_{\bar{q}}}}
(x,y,\boldsymbol{k}_T,\boldsymbol{r}_T),
\end{align}	
where $U$,$U_B$, $u$ and $\nu$ are the Dirac spinors of the proton, and the baryon in the fluctuated state, the quark and the antiquark inside the meson, respectively. $k$ and $r$ denote the momenta of the antiquark and the meson, with $x$ and $y$ representing their light-cone momentum fractions.
The indices $\lambda_N$, $\lambda_B$, $\lambda_q$, $\lambda_{\bar{q}}$ denote the helicity of the proton, the baryon, the quark and sea quark, respectively. 
To evaluate Eq.~(\ref{eq2}), we apply the conventions in Refs.~\cite{Lepage:1980aa, meissner2007relations} for the spinors (in the case of quark and antiquark):
\begin{align}
&u(k,+)=\frac{1}{\sqrt{2^{3/2}k^+}}
\begin{pmatrix}\sqrt{2}k^++m\\k_1+ik_2\\\sqrt{2}k^+-m\\k_1+ik_2\\
\end{pmatrix}, \label{eq3} \\&u(k,-)=\frac{1}{\sqrt{2^{3/2}k^+}}\begin{pmatrix} k_1+ik_2\\\sqrt{2}k^++m\\k_1-ik_2\\
\sqrt{2}k^+-m\\\end{pmatrix},\label{eq4} \\
&\nu(k,+)=\frac{1}{\sqrt{2^{3/2}xr^+}}\begin{pmatrix}\sqrt{2}xr^+-m\\k_1+ik_2\\
\sqrt{2}k^++m\\k_1+ik_2\\ \end{pmatrix},  \label{eq5}\\
&\nu(k,-)=\frac{1}{\sqrt{2^{3/2}k^+}}\begin{pmatrix}-k_1+ik_2\\
\sqrt{2}k^+-m\\k_1-ik_2\\ \sqrt{2}k^+-m\\ \end{pmatrix},  \label{eq6}
\end{align}	
where $m$ is the quark/antiquark mass, $k_i$ is the i-th component of the transverse momenta. 
Similarly, replacing $k$ to $P $ (or $P-r$), and $m$ to $M$ (or $M_B$), one can obtain the spinors of the proton (or baryon $M$). 
In principle there could be higher-spin components for the $\Delta$ baryon, in this work we only consider the spin-$1/2$ components for simplicity.

After some algebra, we find that the wave function $\psi^{\lambda_N}_{{\lambda_B}{\lambda_q}{\lambda_{\bar{q}}}}$ can be organized to the following form
\begin{align}
\psi^{\lambda_N}_{{\lambda_B}{\lambda_q}{\lambda_{\bar{q}}}}(x,y,\boldsymbol{k}_T,\boldsymbol{r}_T)
=&\psi^{\lambda_N}_{\lambda_B}(y,\boldsymbol{r}_T)\psi_{{\lambda_q}{\lambda_{\bar{q}}}}
(x,y,\boldsymbol{k}_T,\boldsymbol{r}_T), \nonumber
\end{align}
where  $\psi^{\lambda_N}_{\lambda_B}(y,r_T)$ can be viewed as the wave function of the nucleon in terms $\pi B$ components, and $\psi_{{\lambda_q}{\lambda_{\bar{q}}}}
(x,y,\boldsymbol{k}_T,\boldsymbol{r}_T)$ is the pion wave function in terms of $q \bar{q}$ components. 
For the former ones, they have the expression:
\begin{align}\label{eq8}
\notag\psi^+_+(y,\boldsymbol{r}_T)&=\frac{M_B-(1-y)M}{\sqrt{1-y}}\phi_1, \displaybreak[0]\\
\notag\psi^+_-(y,\boldsymbol{r}_T)&=\frac{r_1+ir_2}{\sqrt{1-y}}\phi_1, \displaybreak[0]\\
\notag\psi^-_+(y,\boldsymbol{r}_T)&=\frac{r_1-ir_2}{\sqrt{1-y}}\phi_1 , \displaybreak[0]\\
\psi^-_-(y,\boldsymbol{r}_T)&=\frac{(1-y)M-M_B}{\sqrt{1-y}}\phi_1,
\end{align}	
and $\phi_1\equiv\phi_1(y,\boldsymbol{r}_T)$ is the wave function in the momentum space
\begin{align}\label{eq9}
\phi_1(y,\boldsymbol{r}_T)=-\frac{g_1(r^2)\sqrt{y(1-y)}}{\boldsymbol{r}_T^2+L_1^2(m_\pi^2)},
\end{align}
where $m_\pi$ is the mass of $\pi$ meson, $g_1(r^2)$ is  for the coupling of the nucleon-pion-baryon vertex, and
\begin{align}\label{eq10}
L_1^2({m_\pi^2})=yM_B^2+(1-y){m_\pi^2}-y(1-y)M^2.
\end{align}

As for the wave functions of the meson fluctuated from the nucleon in terms of $q\bar{q}$,  $\psi_{{\lambda_q}{\lambda_{\bar{q}}}}(x,y,\boldsymbol{p}_T,\boldsymbol{r}_T)$, they have the following expressions:
\begin{align}\label{eq11}
\notag\psi{_+}{_+}(x,y,\boldsymbol{k}_T,\boldsymbol{r}_T)&=\frac{my}{\sqrt{x(y-x)}}\phi_2,\displaybreak[0]\\
\notag\psi{_+}{_-}(x,y,\boldsymbol{k}_T,\boldsymbol{r}_T)&=
\frac{y(k_1-ik_2)-x(r_1-ir_2)}{\sqrt{x(y-x)}}\phi_2, \displaybreak[0]\\
\notag\psi{_-}{_+}(x,y,\boldsymbol{k}_T,\boldsymbol{r}_T)&=
\frac{y(k_1+ik_2)-x(r_1+ir_2)}{\sqrt{x(y-x)}}\phi_2, \displaybreak[0]\\
\psi{_-}{_-}(x,y,\boldsymbol{k}_T,\boldsymbol{r}_T)&=\frac{-my}{\sqrt{x(y-x)}}\phi_2.
\end{align}
Again $\phi_2=\phi_2(x,y,\boldsymbol{k}_T,\boldsymbol{r}_T)$ is the wave function in the momentum space
\begin{align}\label{eq12}	     	\phi_2(x,y,\boldsymbol{k}_T,\boldsymbol{r}_T)=\frac{g_2\sqrt{\frac{x}{y}(1-\frac{x}{y})}}{(\boldsymbol{k}_T-\frac{x}{y}\boldsymbol{r}_T)^2+L_2^2(m^2)},
\end{align}
$g_2(k^2)$ is the coupling of the pion-quark-antiquark vertex, and
\begin{align}\label{eq13}
L_2^2(m^2)=\frac{x}{y}m^2+\left(1-\frac{x}{y}\right)m^2-\frac{x}{y}\left(1-\frac{x}{y}\right){m_\pi}^2.
\end{align}

For the coupling $g_1$ and $g_2$, we adopt the dipolar form factor
\begin{align}\label{eq14}
g(p^2)=g\frac{p^2-m^2}{|p^2-\Lambda^2_X|^2},
\end{align}
therefore,
\begin{align}
g_1(r^2)&=-g_1(1-y)\frac{\boldsymbol{r}_T^2+L_1^2(m_\pi^2)}
{[\boldsymbol{r}_T^2+L_1^2(\Lambda^2_\pi)]^2},\label{eq15}	\\
g_2(k^2)&=-g_2(1-\frac{x}{y})\frac{(\boldsymbol{k}_T-\frac{x}{y}\boldsymbol{r}_T)^2+L_2^2(m^2)}
{[(\boldsymbol{k}_T-\frac{x}{y}\boldsymbol{r}_T)^2+L_2^2(\Lambda^2_{\bar{q}})]^2}.
\label{eq16}	
\end{align}
Here, $g_1$ and $g_2$ are the free parameters which represent the strength of the nucleon-pion-baryon vertex and pion-quark-antiquark vertex, respectively; and $\Lambda_\pi$ and $\Lambda_{\bar{q}}$ are the cutoff parameters.

\section{Sea quark Sivers function in the overlap representation}

The unpolarized distribution of sea quarks can be calculated from the overlap representation of the proton wave function derived in the previous section~\cite{Brodsky:2000ii}:
\begin{align}\label{eq17}
\notag &f_1(x,y,\boldsymbol{k}_T,\boldsymbol{r}_T)\\
\notag&=\frac{1}{\left(16\pi^3\right)^2}\frac{1}{2}\sum_{\{\lambda\}}
|\psi^{\lambda_N}_{{\lambda_B}{\lambda_q}{\lambda_{\bar{q}}}}(x,y,\boldsymbol{k}_T,\boldsymbol{r}_T)|^2\\
\notag&=\frac{1}{\left(16\pi^3\right)^2}\frac{1}{2}\sum_{\lambda_N, \lambda_B}
|\psi^{\lambda_N}_{\lambda_B}(y,\boldsymbol{r}_T)|^2\sum_{{\lambda_q}{\lambda_{\bar{q}}}}
|\psi_{{\lambda_q}{\lambda_{\bar{q}}}}(x,y,\boldsymbol{k}_T,\boldsymbol{r}_T)|^2\\
&=f_{1}^{\pi/p}(y,\boldsymbol{r}_T)
f_{1}^{\bar{q}/\pi}\left(\frac{x}{y},\boldsymbol{k}_T-\frac{x}{y}\boldsymbol{r}_T\right),
\end{align}		
where $\{\lambda\}={\lambda_N,\lambda_B, \lambda_q, \lambda_{\bar{q}}}$. In the above equation,  $f_{1}^{\pi/p}$ denotes the distribution of the pion inside the proton:
\begin{align}\label{eq13}
\notag f_{1}^{\pi/P}(y,\boldsymbol{r}_T)&=\frac{1}{16\pi^3}(|\psi^+_{+}|^2+|\psi^+_{-}|^2)\\
&=\frac{g_1^2}{16\pi^3}\frac{y(1-y)^2[\boldsymbol{r}^2_T+(M_B-M(1-y))^2]}
{[\boldsymbol{r}_T^2+L_1^2(\Lambda^2_\pi)]^4}
\end{align}
while $f_{1}^{\bar{q}/\pi}(\frac{x}{y},\boldsymbol{k}_T-\frac{x}{y}\boldsymbol{r}_T)$ is the sea quark unpolarized TMD distribution in a pion with the momentum fraction $x/y$ and the intrinsic transverse momentum $\boldsymbol{k}_T-\frac{x}{y}\boldsymbol{r}_T$:
\begin{align}\label{eq20}
\notag&f_{1}^{\bar{q}/\pi}(\frac{x}{y},\boldsymbol{k}_T-\frac{x}{y}\boldsymbol{r}_T)\\
\notag&=\frac{1}{16\pi^3}(|\psi{_+}{_+}|^2+|\psi{_+}{_-}|^2\notag+|\psi{_-}{_+}|^2+|\psi{_-}{_-}|^2)\\
&=\frac{g^2_2}{8\pi^3}\frac{(1-\frac{x}{y})^2[m^2+(\boldsymbol{k}_T-\frac{x}{y}\boldsymbol{r}_T)^2]}
{[(\boldsymbol{k}_T-\frac{x}{y}\boldsymbol{r}_T)^2+L^2_2(\Lambda^2_{\bar{q}})]^4}.
\end{align}

After integrating out the longitudinal momentum fraction $y$ and the transverse momentum $\boldsymbol{r}_T$, we can obtain the unpolarized distribution of the sea quarks inside the proton
\begin{align}\label{eq18}
\notag&f_{1}^{\bar{q}/p}(x,\boldsymbol{k}_T)\\
&=\int_{x}^{1}\frac{dy}{y}\int{d^2\boldsymbol{r}_T}
f_{1}^{\pi/p}(y,\boldsymbol{r}_T)f_{1}^{\bar{q}/\pi}(\frac{x}{y},\boldsymbol{k}_T-\frac{x}{y}\boldsymbol{r}_T).
\end{align}			
Thus, having the LCWFs of the proton in terms of $|\bar{q}q B\rangle$, the unpolarized sea quark distribution in the proton can be expressed as the convolution of the $\pi$ distribution in the proton $f_{1}^{\pi/p}(y,\boldsymbol{r}_T)$ and the antiquark distribution in the pion $f_{1}^{\bar{q}/\pi}(\frac{x}{y},\boldsymbol{k}_T-\frac{x}{y}\boldsymbol{r}_T)$.

Using the overlap representation of the LCWFs, the sea quarks Sivers function can be calculated from the following expression~\cite{Lu:2006kt,Bacchetta:2008af}
\begin{align}	
&\frac{2(\widehat{\boldsymbol{s}}_T\times\boldsymbol{k}_T)\cdot\widehat{\boldsymbol{P}}}
{M}f_{1T}^\bot(x,y,\boldsymbol{k}_T^2,\boldsymbol{r}_T^2)
=\int\frac{d^2\boldsymbol{r}^{\prime}_T}{16\pi^3}G(x,\boldsymbol{k}_T,\boldsymbol{k}^{\prime}_T)\nonumber\\
&\sum_{\{\lambda\}}
[\psi^{\uparrow*}_{{\lambda_B}{\lambda_q}{\lambda_{\bar{q}}}}
(x,y,\boldsymbol{k}_T,\boldsymbol{r}_T)\psi^{\uparrow}_{{\lambda_B}{\lambda_q}
{\lambda_{\bar{q}}}}(x,y,\boldsymbol{k}_T,\boldsymbol{r}^{\prime}_T)\nonumber\\
&-\psi^{\downarrow*}_{{\lambda_B}{\lambda_q}{\lambda_{\bar{q}}}}
(x,y,\boldsymbol{k}_T,\boldsymbol{r}_T)\psi^{\downarrow}_{{\lambda_B}{\lambda_q}
{\lambda_{\bar{q}}}}(x,y,\boldsymbol{k}_T,\boldsymbol{r}^{\prime}_T)] \label{eq:sivers}
\end{align}
Here, $\uparrow$ and $\downarrow$ denote the transverse polarization states of proton. The function $G(x,\boldsymbol{k}_T,\boldsymbol{k}^{\prime}_T)$ is the factor simulating the final state interaction (in DIS) between the active antiquark and the spectator~\cite{Lu:2006kt,Bacchetta:2008af},
In this work we adopt the following form
\begin{align}\label{eq:ImG}
\textrm{Im}\, G(x,\boldsymbol{k}_T,\boldsymbol{k}^{\prime}_T)
=-\frac{C_F\alpha_s}{2\pi}\frac{1}{(\boldsymbol{k}_T-\boldsymbol{k}^{\prime}_T)^2},
\end{align}
corresponding to the one-gluon exchange approximation of the gauge-link.

Substituting the light-cone wave functions of the proton into Eq.(\ref{eq:sivers}), we find that the sea quark sivers function in our model has the form:
\begin{widetext}
\begin{align}\label{eq25}       	
\notag f_{1T}^{\bot,\bar{q}/p}(x,\boldsymbol{k}^2_T)&=-\int_{x}^{1}\frac{dy}{y}\int{d^2\boldsymbol{r}_T}
\int{d^2\boldsymbol{l}_T}\frac{g_1^2C_F\alpha_s}{16\pi^4}\frac{\boldsymbol{l}_T\cdot\boldsymbol{k}_T}
{\boldsymbol{k}_T^2}  \frac{My(1-y)^2[M_B-M(1-y)]}{{L_1^2(\Lambda^2_\pi)} [\boldsymbol{r}_T^2+L_1^2(\Lambda^2_\pi)]^2[(\boldsymbol{r}_T-\boldsymbol{l}_T)^2+L_1^2(\Lambda^2_\pi)]^2}
\\&\times\frac{g^2_2}{8\pi^3}\frac{(1-\frac{x}{y})^2[m^2+(\boldsymbol{k}_T-\frac{x}{y}\boldsymbol{r}_T)^2]}
{[(\boldsymbol{k}_T-\frac{x}{y}\boldsymbol{r}_T)^2+L^2_2(\Lambda^2_{\bar{q}})]^2
[(\boldsymbol{k}_T-\frac{x}{y}\boldsymbol{r}_T-(1-\frac{x}{y})\boldsymbol{l}_T)^2
+L^2_2(\Lambda^2_{\bar{q}})]^2}.
\end{align}
\end{widetext}
with $\bm l_T= \bm k_T-\bm k_T^\prime=\bm r_T-\bm r_T^\prime$.
The integration over $y$, $\bm r_T$ and $\bm l_T$ can be performed numerically.

In the above calculations we have adopted the dipolar form factor for the pion-quark-antiquark coupling in  $\psi(x,y,\bm k_T^\prime, \bm r_T^\prime)$ which explicitly depends on the transverse momentum flow $\bm l_T$:
\begin{align}
g_2(k^{\prime ^2})&=-g_2(1-\frac{x}{y})\frac{((\boldsymbol{k}_T-\frac{x}{y}\boldsymbol{r}_T-(1-\frac{x}{y})\boldsymbol{l}_T)^2
+L_2^2(m^2)}
{(\boldsymbol{k}_T-\frac{x}{y}\boldsymbol{r}_T-(1-\frac{x}{y})\boldsymbol{l}_T)^2
+L^2_2(\Lambda^2_{\bar{q}})^2}.\label{eq:form_factor2}	
\end{align}

\begin{center}\label{tab1}
	\setlength{\tabcolsep}{5mm}
	\renewcommand\arraystretch{1.5}
	\begin{tabular}{ c | c | c }
		\hline
		Parameters & $\bar{u}$ & $\bar{d}$ \\
		\hline
		\hline
		$g_1$ & 9.33 & 5.79 \\
		\hline
		$g_2$ & 4.46 & 4.46 \\
		\hline
		$\Lambda_\pi (\textrm{GeV})$ &  0.223 & 0.223 \\
		\hline
		$\Lambda_{\bar{q}} (\textrm{GeV})$ &  0.510 &  0.510 \\
		\hline
	\end{tabular}
	\captionof{table}{Values of the parameters obtained from fitting the model to the GRV LO (second and forth row) and MSTW2008 (first and third row).} \label{tab1}
\end{center}

\begin{figure*}[htbp]
       			\includegraphics[width=0.45\linewidth]{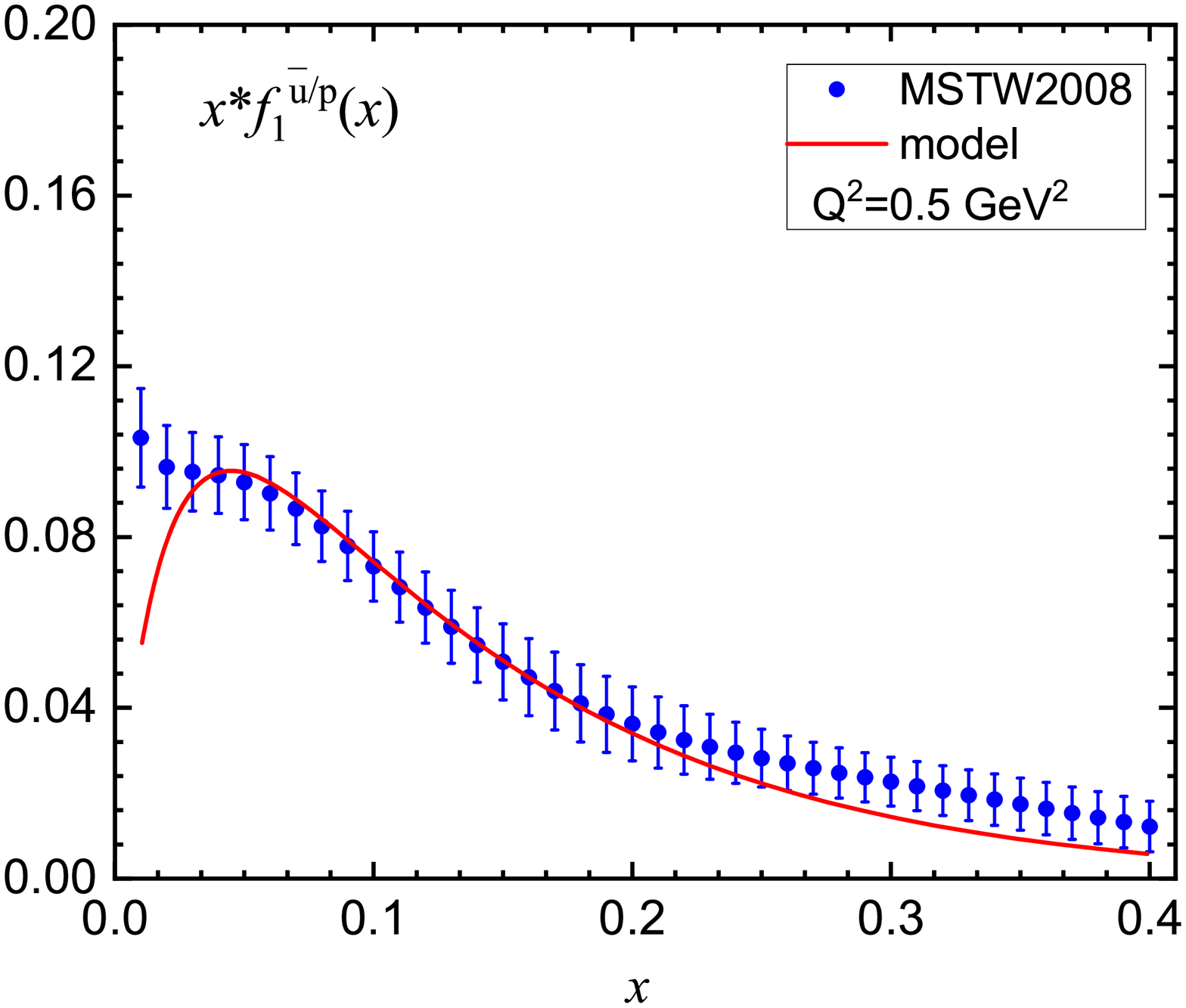}
       			\includegraphics[width=0.45\linewidth]{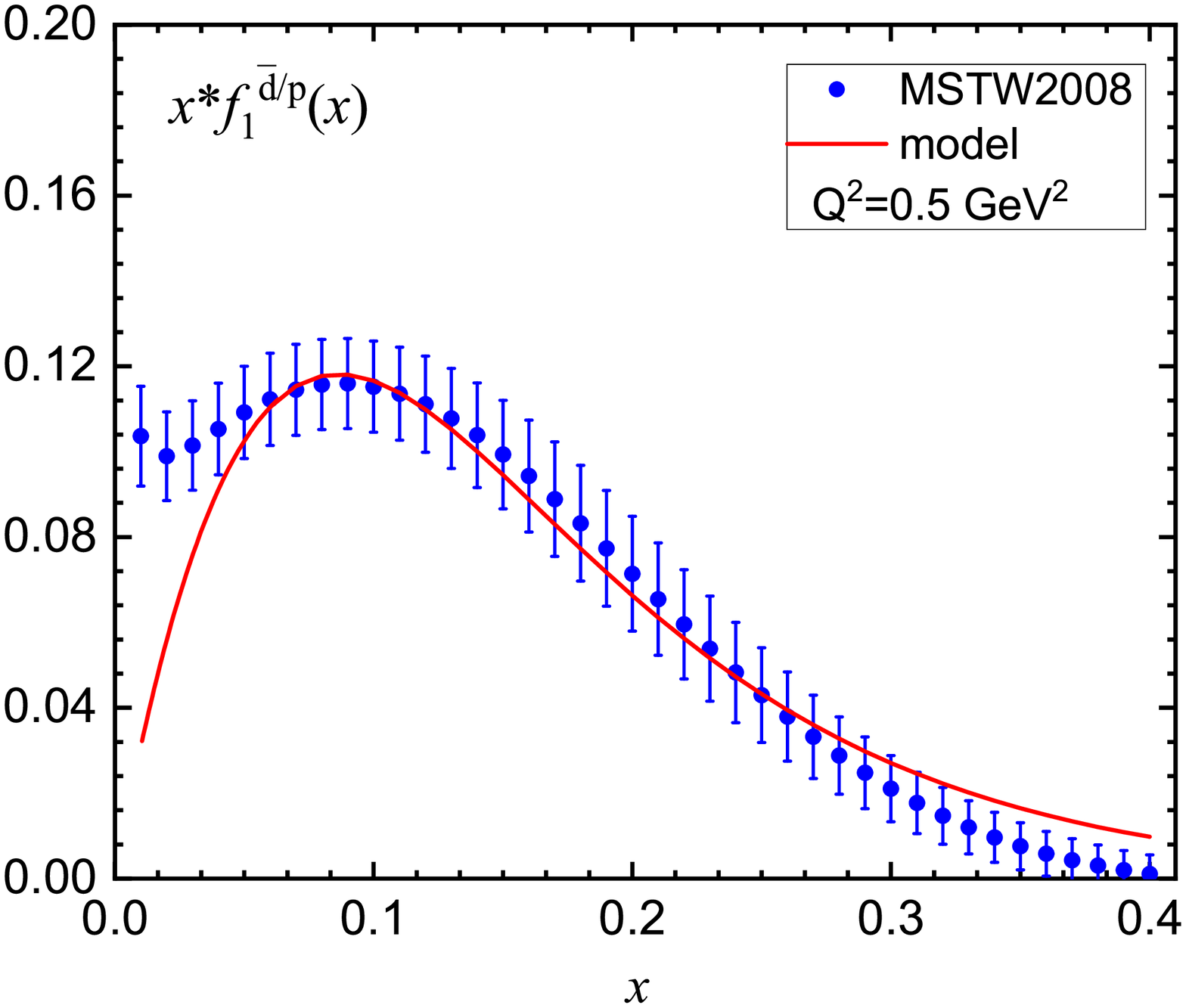}	
       	\caption{The fitting of the model results to the MSTW2008 LO parametrization for the sea quark distribution $f_{1}^{\bar{q}/p}(x)$ multiplied by $x$ at $Q^2=0.5$ GeV$^2$. The left panel shows the result of $\bar{u}$ and the right shows the result of $\bar{d}$. The error bars corresponds to the uncertainty of the MSTW2008 parametrization at the 90$\%$ C.L.} \label{f1}
\end{figure*}

\begin{figure*}[htbp]
       	\centering
       	\subfigure{\begin{minipage}[b]{0.45\linewidth}
       			\centering
       			\includegraphics[width=0.95\linewidth]{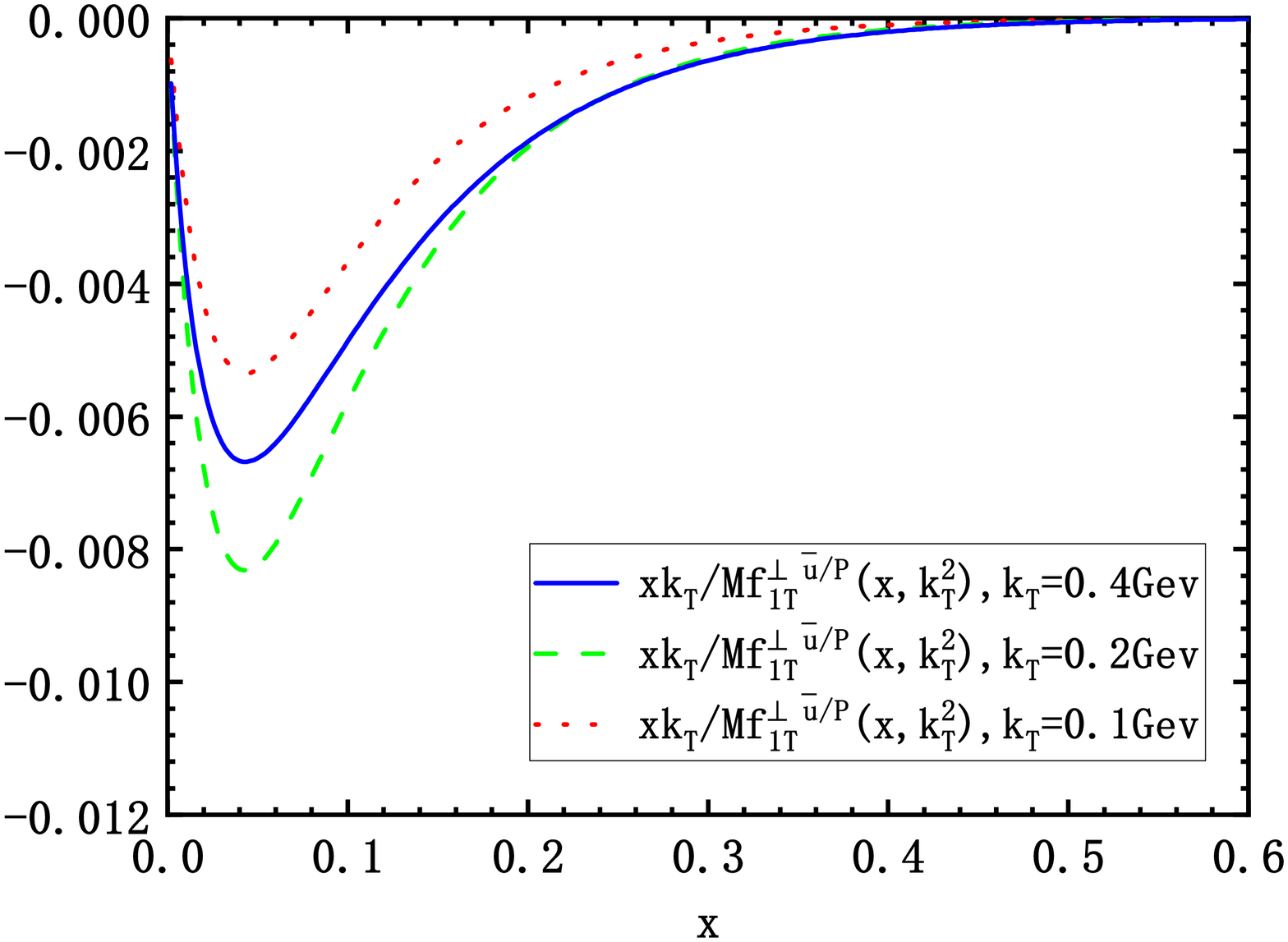}
       	\end{minipage}}
       	\subfigure{\begin{minipage}[b]{0.45\linewidth}
       			\centering
       			\includegraphics[width=0.95\linewidth]{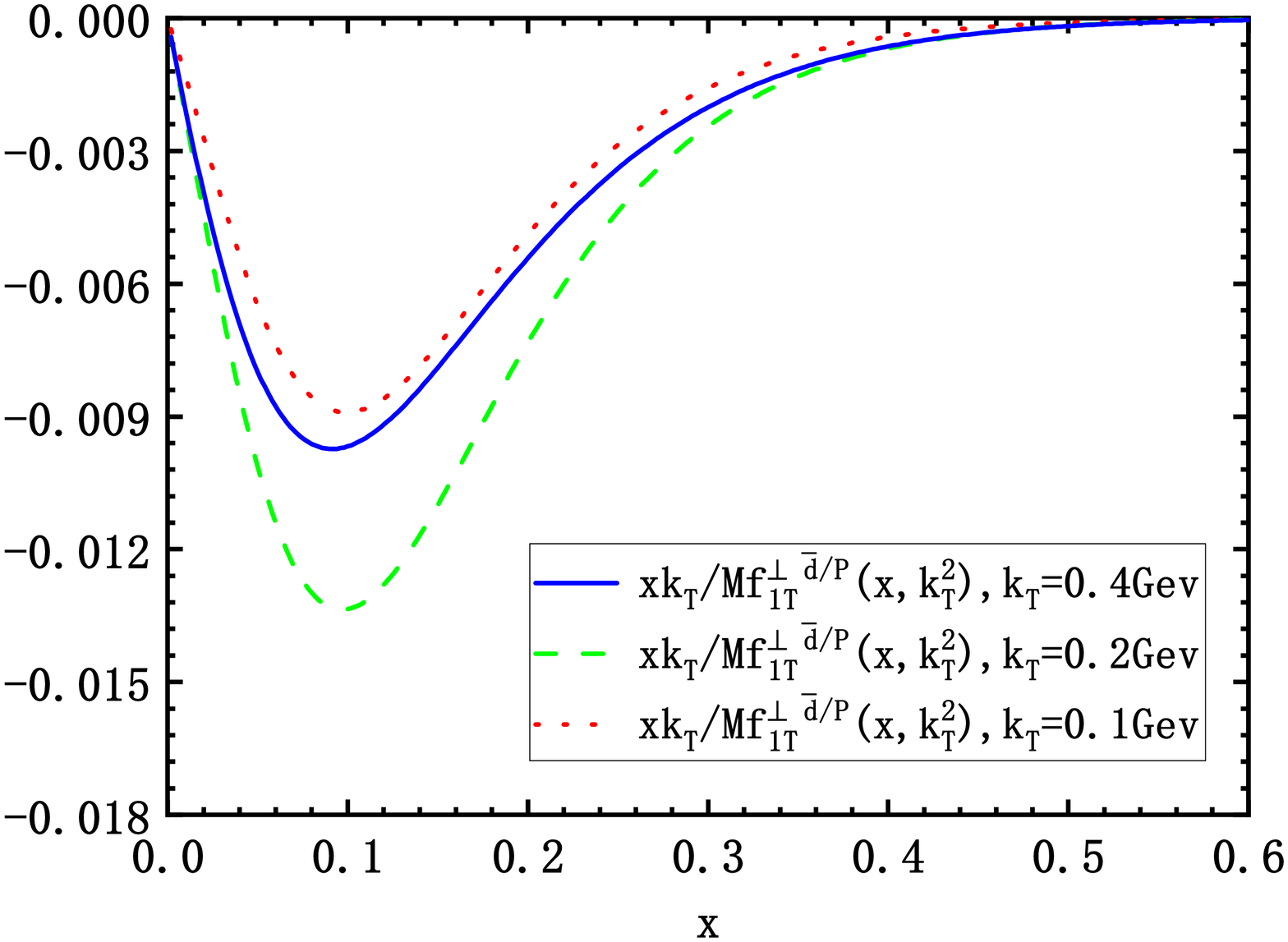}	
       	\end{minipage}}
       	\caption{ The $x$-dependence of $x\frac{k_T}{M}	f_{1T}^{\bot,\bar{u}/p}(x,\boldsymbol{k}^2_T)$ (left panel) and $x\frac{k_T}{M}	f_{1T}^{\bot,\bar{u}/p}(x,\boldsymbol{k}^2_T)$ (right panel) at the fixed transverse momenta $k_T=0.1$ GeV, 0.2 GeV, 0.4 GeV, respectively.}\label{f2}
\end{figure*}

In a simplified calculation, we assume that the form factor $g_2(k^{\prime ^2})$ does not depend on $\bm l_T$, that is, $g_2(k^{\prime^2})\rightarrow g_2(k^2)$, which has been suggested in several model calculations~\cite{Bacchetta:2002tk,bacchetta2004sivers,Amrath:2005gv,Bacchetta:2007wc}.
In doing this, it indicates the assumption that the final-state interaction occurs between the pion meson as a whole and the spectator baryon, which has been adopted in the chiral model calculation~\cite{he2019sivers}.
In this case, we find that the integration over $\bm l_T$ can be performed analytically and Eq.~(\ref{eq:sivers}) yields the following result
\begin{align}\label{eq26}	
	\notag &k^i_Tf_{1T}^\bot(x,y,\boldsymbol{k}^2_T,\boldsymbol{r}^2_T)	\\&=16\pi^3r^i_Tf_{1T}^{\bot,\pi/p}(y,\boldsymbol{r}^2_T)
f_{1}^{\bar{q}/\pi}(\frac{x}{y},\boldsymbol{k}_T-\frac{x}{y}\boldsymbol{r}_T).
\end{align}
Here, $f_{1T}^{\bot,\pi/p}(y,\boldsymbol{r}^2_T)$ is defined as the ``Sivers function of the pion" inside the proton~\cite{he2019sivers}, and in our model it may be calculated from the light-cone wave functions of the proton in terms of $\pi B$ component,
\begin{align}\label{eq28}
\notag&\frac{2(\widehat{\boldsymbol{s}}_T\times\boldsymbol{r}_T)\cdot
\widehat{\boldsymbol{P}}}{M}f_{1T}^{\bot,\pi/P}(y,\boldsymbol{r}^2_T)
=\int\frac{d^2\boldsymbol{r}^{\prime}_T}{16\pi^3}G(y,\boldsymbol{r}_T,\boldsymbol{r}^{\prime}_T) \\&\sum_{\lambda_B}[\psi^{\uparrow*}_{\lambda_B}(y,\boldsymbol{r}_T)
\psi^{\uparrow}_{\lambda_B}(y,\boldsymbol{r}^{\prime}_T)
-\psi^{\downarrow*}_{\lambda_B}(y,\boldsymbol{r}_T)
\psi^{\downarrow}_{\lambda_B}(y,\boldsymbol{r}^{\prime}_T)].
\end{align}
where $\textrm{Im} G(y,\boldsymbol{r}_T,\boldsymbol{r}^{\prime}_T)$ has the same form of Eq.~(\ref{eq:ImG}), which simulates the final state interaction between the pion meson and the spectator baryon.
Furthermore, is is easy to obtain the analytical result of $f_{1T}^{\bot,\pi/P}(y,\boldsymbol{r}^2_T)$ from the above equation:
\begin{align}\label{eq22}
\notag&f_{1T}^{\bot,\pi/p}(y,\boldsymbol{r}^2_T)=-\frac{g_1^2C_F\alpha_s}{4\pi^2} \frac{My(1-y)^2[M_B-M(1-y)]}{[\boldsymbol{r}_T^2+L_1^2(\Lambda^2_\pi)]^2} \\&\times\int\frac{d^2\boldsymbol{r}^{\prime}_T}{(2\pi)^2}  \frac{(\boldsymbol{r}_T-\boldsymbol{r}^{\prime}_T)\cdot\boldsymbol{r}_T}{\boldsymbol{r}^2_T}  \frac{1}{(\boldsymbol{r}_T-\boldsymbol{r}^{\prime}_T)^2[\boldsymbol{r}^{\prime}_T+L_1^2(\Lambda^2_\pi)]^2}.
\end{align}

Thus, similar to the case of unpolarized distribution, in this simplified scenario, the sea quark Sivers function can be expressed as the convolution of $f_{1T}^{\bot,\pi/P}(y,\boldsymbol{r}^2_T)$ and $f_{1}^{\bar{q}/\pi}(x,\boldsymbol{k}^2_T)$,
\begin{align}\label{eq33}	f_{1T}^{\bot,\bar{q}/p}(x,\boldsymbol{k}^2_T)&=-\int_{x}^{1}\frac{dy}{y}\int{d^2\boldsymbol{r}_T}\frac{\boldsymbol{r}_T\cdot\boldsymbol{k}_T}{\boldsymbol{k}_T^2} \notag	\\&\notag\times\frac{g^2_2}{8\pi^3}\frac{(1-\frac{x}{y})^2[m^2+(\boldsymbol{k}_T-\frac{x}{y}\boldsymbol{r}_T)^2]}{[(\boldsymbol{k}_T-\frac{x}{y}\boldsymbol{r}_T)^2+L^2_2(\Lambda^2_{\bar{q}})]^4}
	\\&\times\frac{g_1^2C_F\alpha_s}{16\pi^3}  \frac{My(1-y)^2[M_B-M(1-y)]}{{L_1^2(\Lambda^2_\pi)} [\boldsymbol{r}_T^2+L_1^2(\Lambda^2_\pi)]^3}.
\end{align}	
The integration over $y$ and $\bm r_T$ can be performed numerically.

\begin{figure*}[htbp]
       	\centering
       	\subfigure{\begin{minipage}[b]{0.45\linewidth}
       			\centering
       			\includegraphics[width=0.95\linewidth]{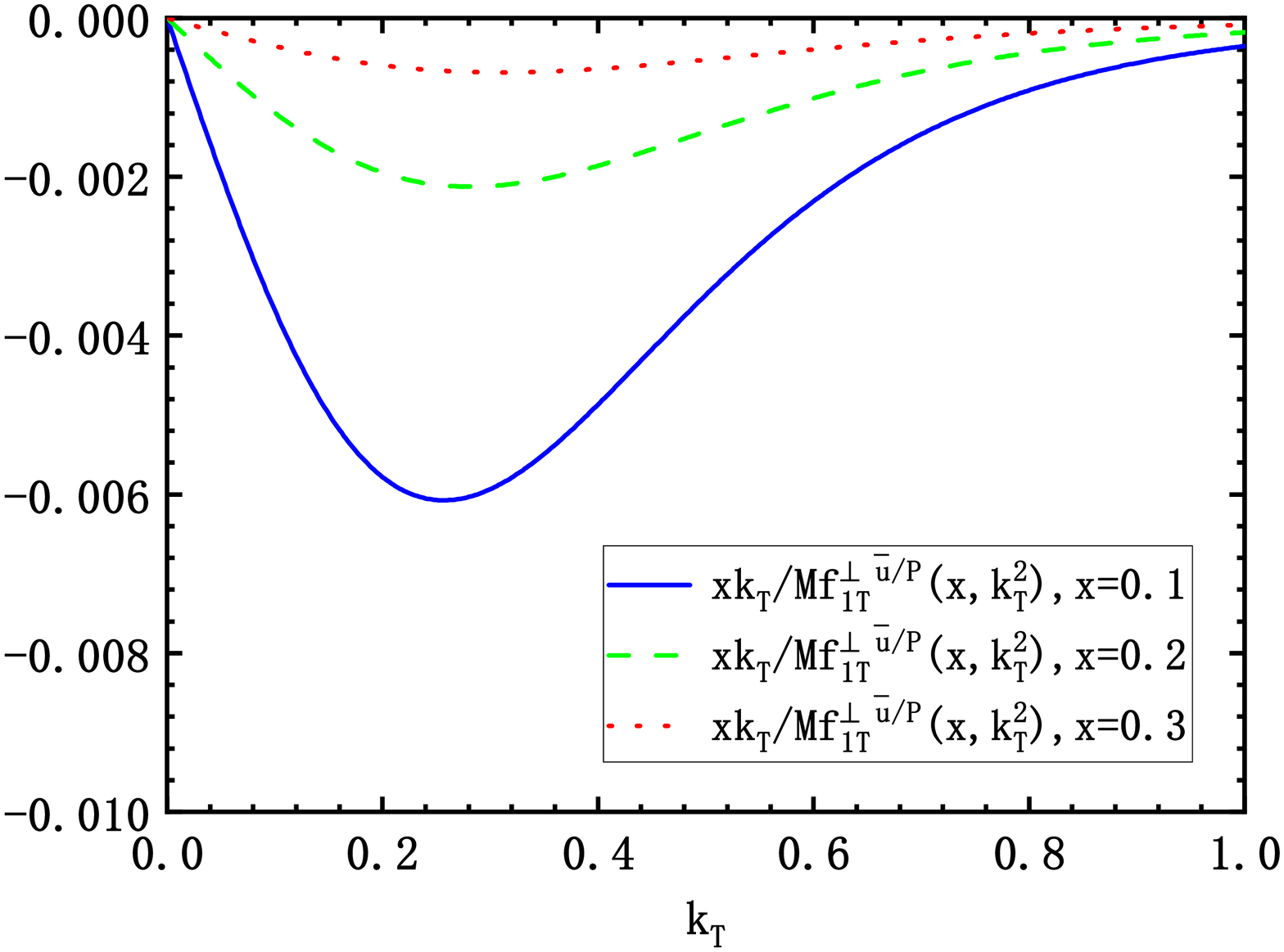}
       	\end{minipage}}
       	\subfigure{\begin{minipage}[b]{0.45\linewidth}
       			\centering
       			\includegraphics[width=0.95\linewidth]{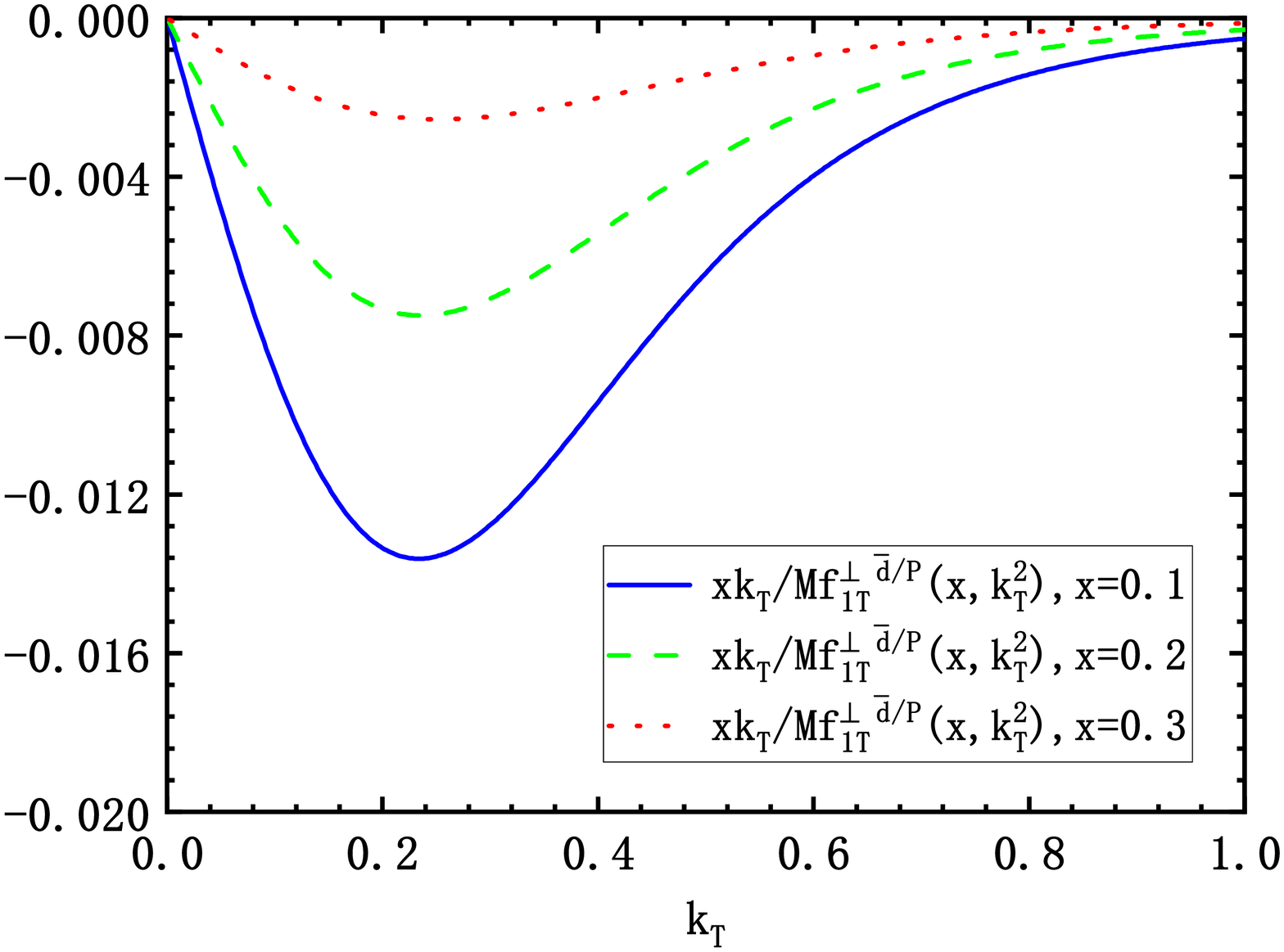}	
       	\end{minipage}}
       	\caption{ Similar to Fig~\ref{f2}, but for the $k_T$-dependence at the fixed momentum fractions $x=0.05$, $x=0.1$, $x=0.2$.}\label{f3}
\end{figure*}

\section{ Numerical Results for the sea quark Sivers function}
\label{Sec4}

In this section, we present the numerical result for the Sivers function of the sea quarks $\bar{u}$ and $\bar{d}$.
To specify the values of the parameters $g_1$, $g_2$, $\Lambda_{\bar{q}}$, $\Lambda_\pi$ in our model, we try to fit the model result to the existing parameterizations for the distributions $f_{1}^{\bar{q}/\pi}(x)$ and $f_{1}^{\bar{q}/p}(x)$, since these unpolarized distributions have been extracted from experimental data and are well known.
In our work, we adopt the GRV leading-order (LO) parametrization~\cite{gluck1992pionic} to perform the fit for $f_{1}^{\bar{u}/\pi^-}(x)$ (or $f_{1}^{\bar{d}/\pi^+}(x)$) as well as adopt the MSTW2008 LO parametrization~\cite{martin2009parton} for $f_{1}^{\bar{u}/p}$ and $f_{1}^{\bar{d}/p}$.
In the first fit, we can obtain the values of the parameters $g_2$ and $\Lambda_{\bar{q}}$;
while in the second fit of $f_{1}^{\bar{q}/p}$,  we can obtain the values of the parameters $g_1$ and   $\Lambda_\pi$. In both fits, we fix the mass for the quark and the sea quark to be $m=0.3~\text{GeV}$.

When doing this, we have to also choose an appropriate energy scale $Q^2$ at which our model can be compared to the parametrization. In the process of fitting, we find $Q^2=0.5$ GeV$^2$ can achieve a fairly good fit.
In Fig.\ref{f1}, we plot the fitted the unpolarized distribution of the sea quarks $f_{1}^{\bar{q}/p}(x)$ (multiplied by $x$) in our model and compare it with the MSTW2008 LO parametrization (solid line). The left panel shows the result of $\bar{u}$ and the right shows the result of $\bar{d}$.
The error bars correspond to the uncertainties of parametrization the at the $90\%$ C.L.
The best values for the parameters from the fits are shown in Table.~\ref{tab1}.

\begin{figure*}[htbp]
       	\centering
       	\subfigure{\begin{minipage}[b]{0.45\linewidth}
       			\centering
       			\includegraphics[width=0.95\linewidth]{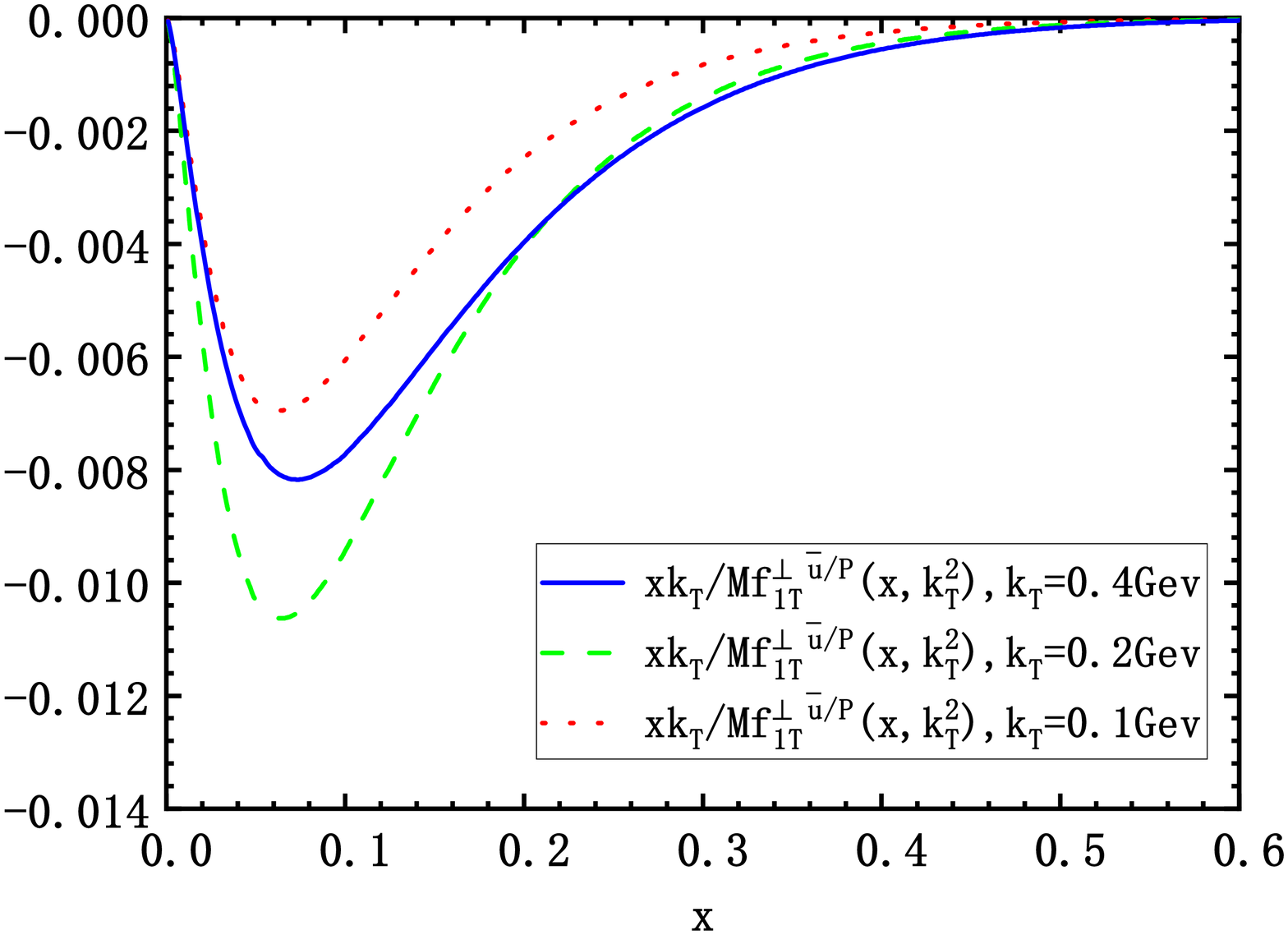}
       	\end{minipage}}
       	\subfigure{\begin{minipage}[b]{0.45\linewidth}
       			\centering
       			\includegraphics[width=0.95\linewidth]{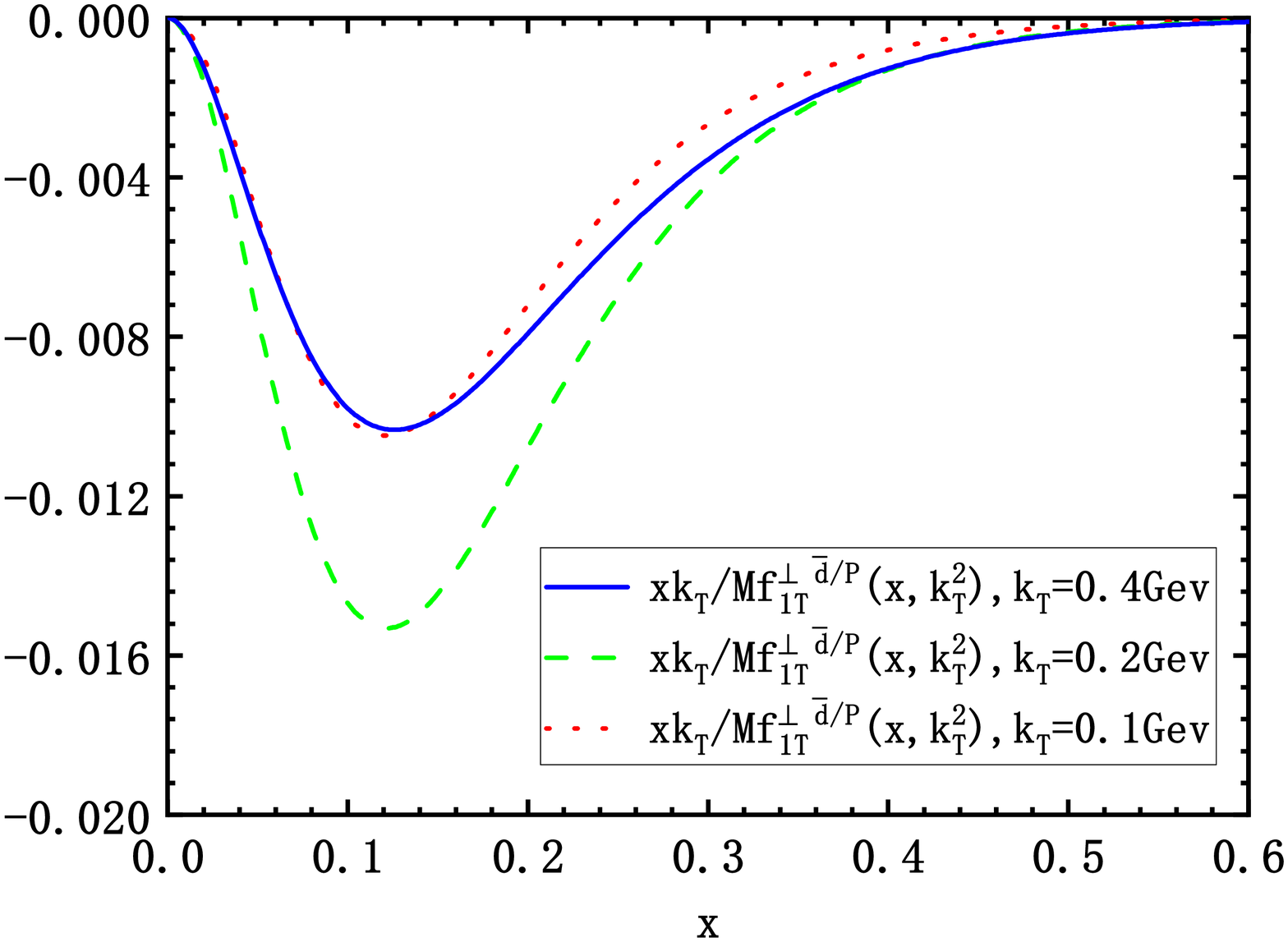}	
       	\end{minipage}}
       	\centering
       	\subfigure{\begin{minipage}[b]{0.45\linewidth}
       			\centering
       			\includegraphics[width=0.95\linewidth]{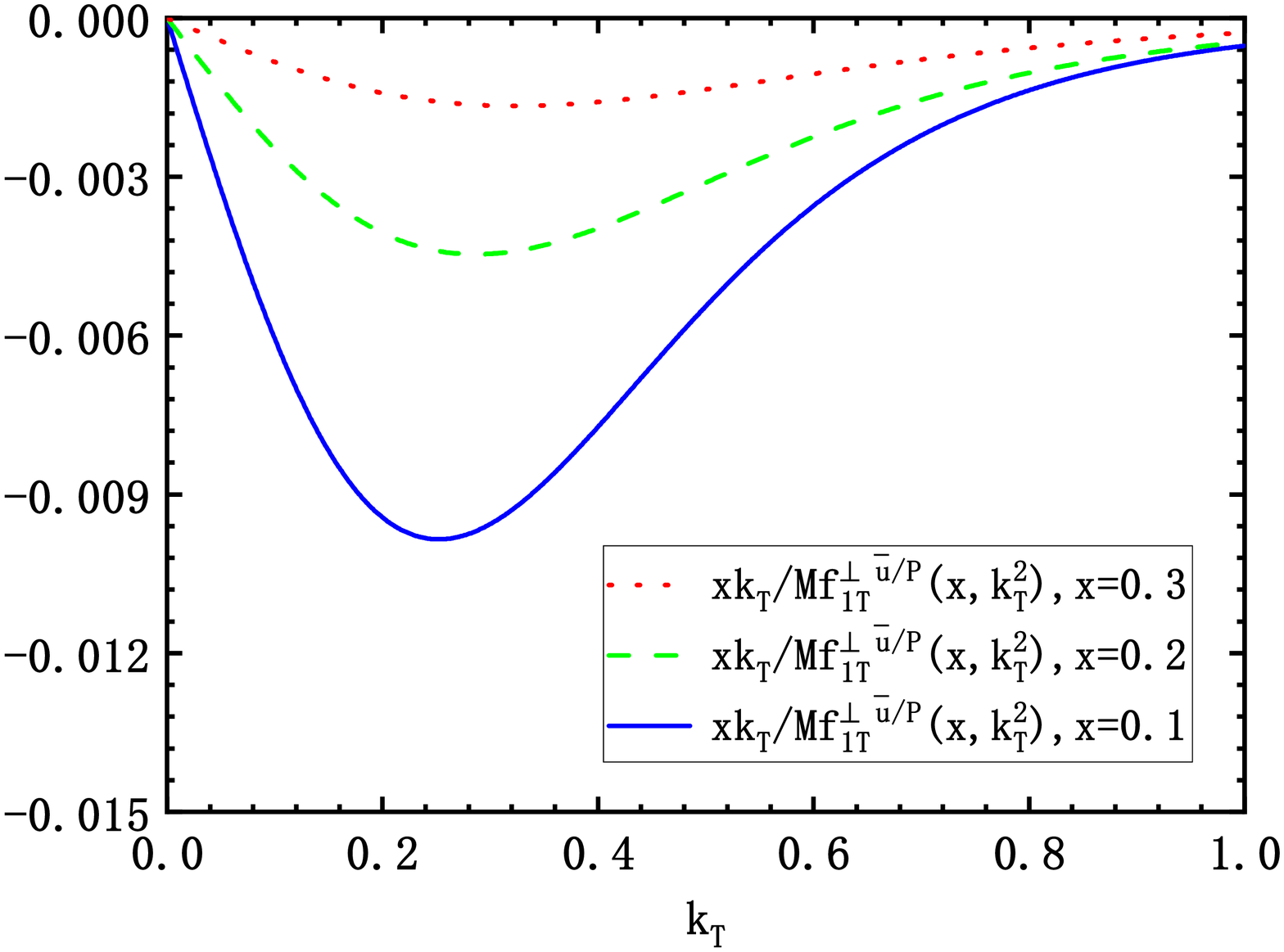}
       	\end{minipage}}
       	\subfigure{\begin{minipage}[b]{0.45\linewidth}
       			\centering
       			\includegraphics[width=0.95\linewidth]{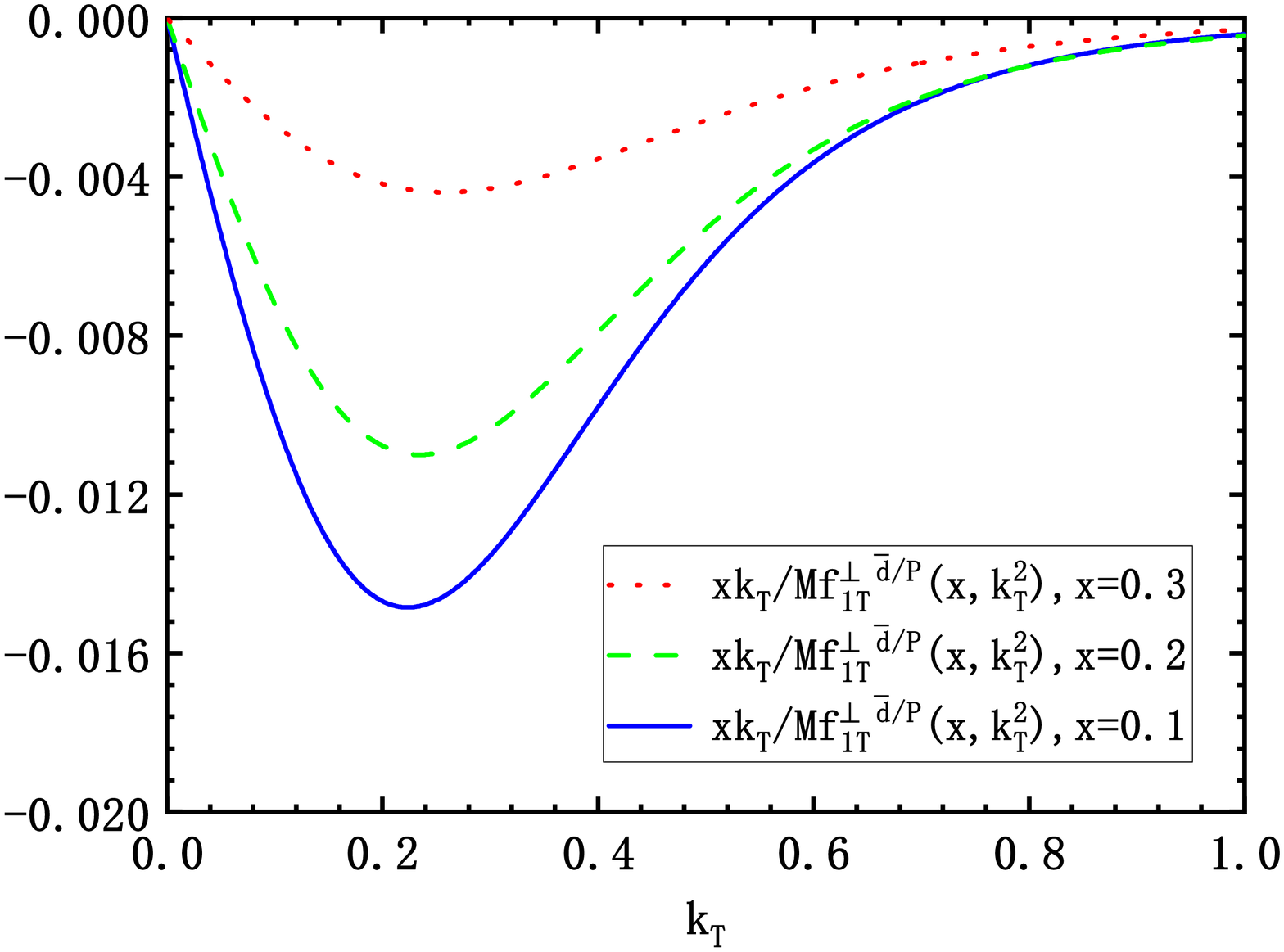}	
       	\end{minipage}}
       	\caption{ Similar to Figs.~\ref{f2} and \ref{f3}, but using the convoluted form in Eq.~\ref{eq33}. }\label{fig:fconv}
\end{figure*}

Using the values of the parameters given in Table.~\ref{tab1}, we calculate the numerical result of the sea Sivers functions at the model scale. 
First we apply the expression in Eq.~(\ref{eq25}) to perform the calculation.
In the left and right panels of Fig.~\ref{f2}, we plot the $x$-dependence of the sea Sivers functions (multiplied by a prefactor $x\frac{k_T}{M}$) of $\bar{u}$ and $\bar{d}$ quarks at the fixed transverse momentum $k_T=0.1$ GeV, $k_T=0.2$ GeV, $k_T=0.4$ GeV, respectively.
In Fig.~\ref{f3}, we also plot the $k_T$-dependence of these functions at the fixed the momentum fraction $x=0.1$, $x=0.2$, $x=0.3$, respectively.
We find that in both the cases of $\bar{u}$ and $\bar{d}$, the signs are negative in the entire $x$ and $k_T$ region. The size of $f_{1T}^{\perp,\bar{d}/p}(x,\boldsymbol{p}^2_T)$ is larger than that of $f_{1T}^{\perp,\bar{u}/p}(x,\boldsymbol{p}^2_T)$ .
Similarly, the $k_T$-dependence of the sea quark TMD distribution function varies with $x$. 
To be specific, as $x$ increases, the peak of the curves shift from lower $k_T$ to higher $k_T$. While as $k_T$ increases, the peak of the curves shift from higher $x$ to lower $x$.

As a comparison, we also apply the convoluted expression in Eq.~(\ref{eq33}) to calculate the sea quark Sivers function. the results are plotted in Fig.~\ref{fig:fconv}, with the upper and lower panels showing the $x$-dependence and the $k_T$-dependence of the function.
We find that these results are consistent with the exact results in Figs.~\ref{f2} and \ref{f3} in sign and size.
This means that the assumption that the final-state interaction occurs between the pion meson and the spectator approximately holds.

\begin{figure*}
   	\centering
  \includegraphics[width=0.9\columnwidth]{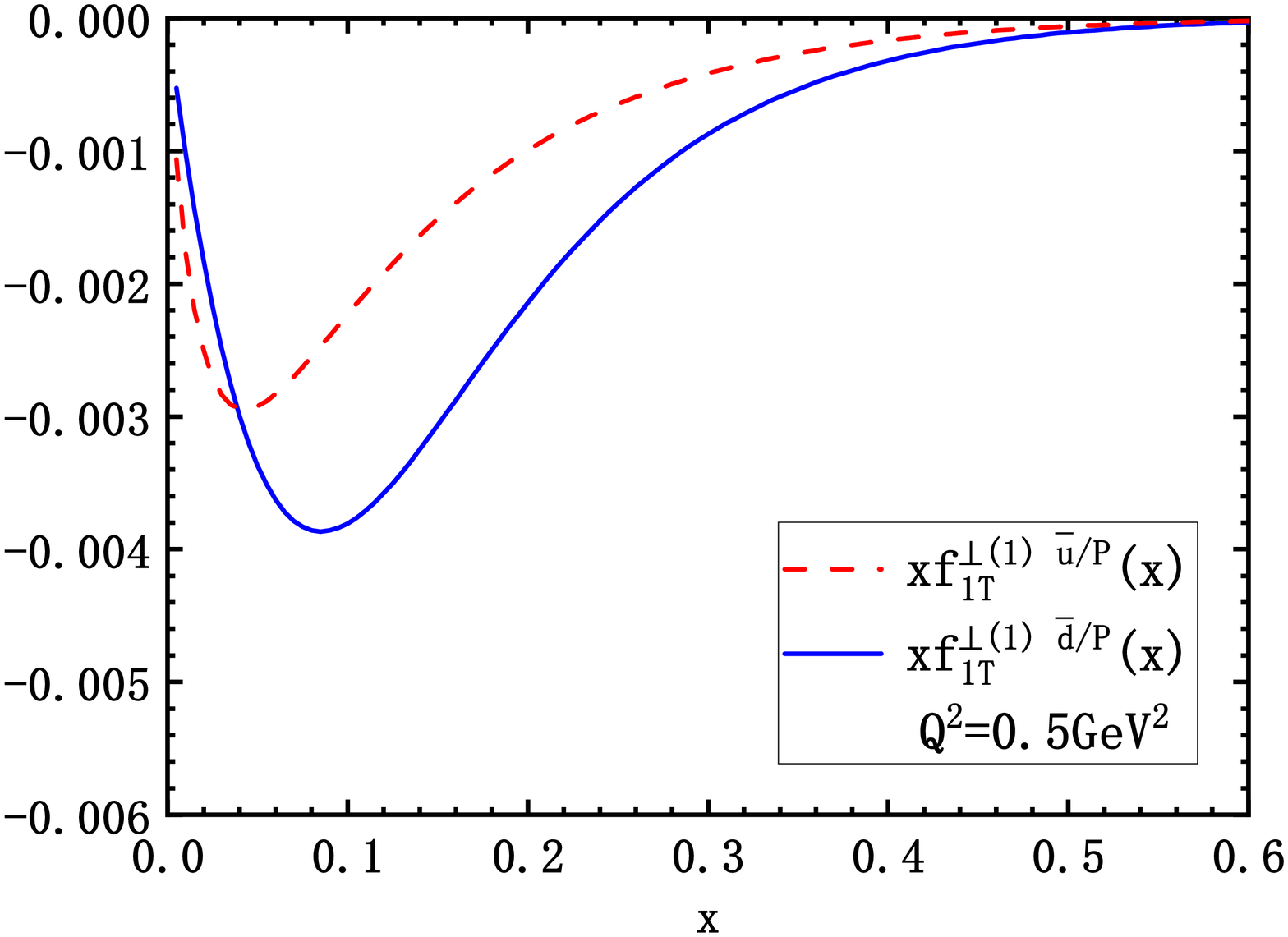}
    \includegraphics[width=0.9\columnwidth]{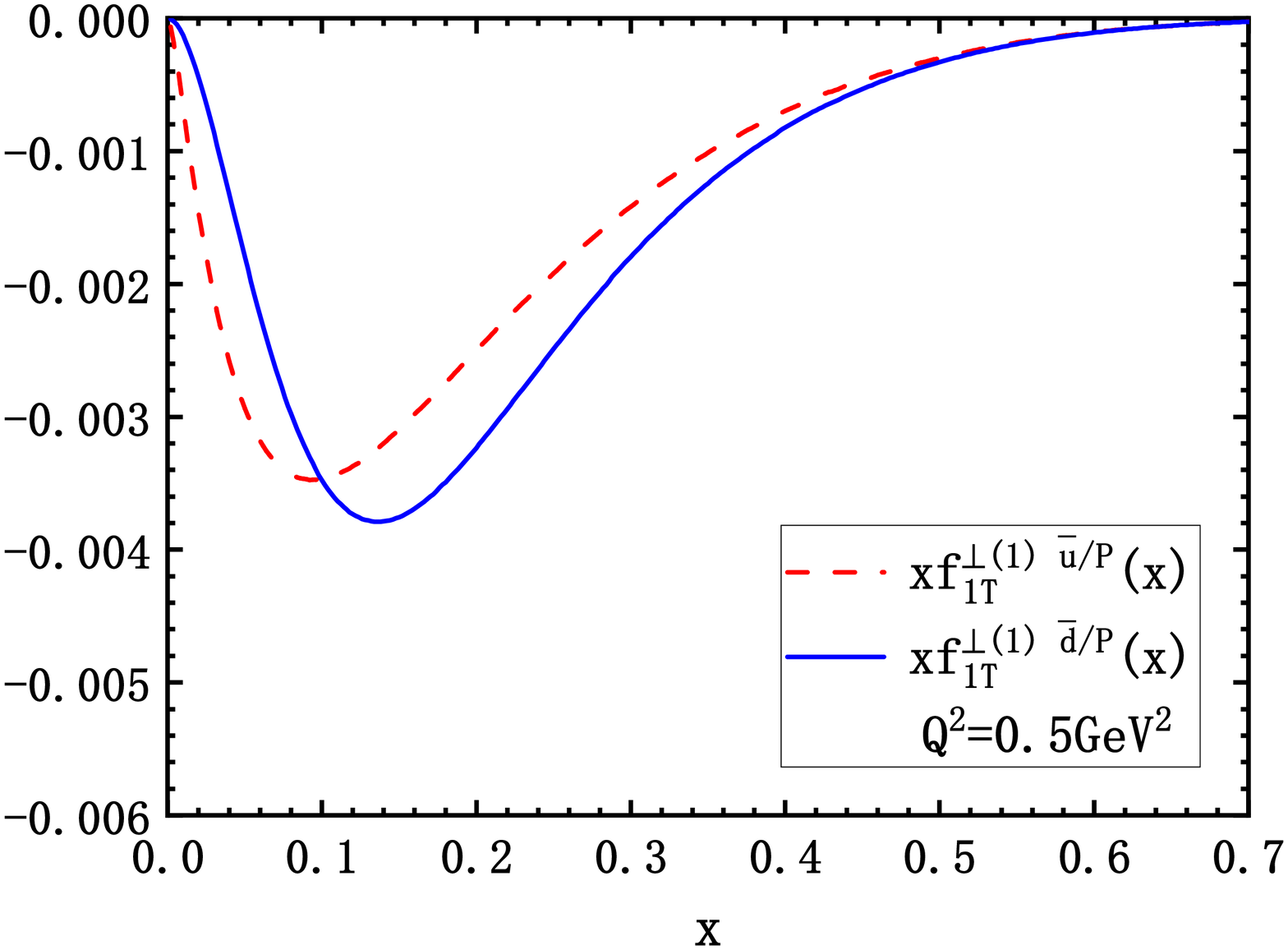}
   	\caption{The first transverse-moment of the  quark Sivers function (timed with $x$) of $\bar{u}$ and $\bar{d}$ quarks at the model scale $Q^2=0.5$ GeV$^2$.}\label{f4}
\end{figure*}

Furthermore, we calculate the first transverse-moment ($k_T$-moment) of the sea quark Sivers function, which is defined as
\begin{align}\label{eq34}	
\notag	
f_{1T}^{\bot(1),\bar{q}/p}(x)&
=\int{d^2\boldsymbol{k}_T}\frac{\boldsymbol{k}_T^2}{2M^2}f_{1T}^{\bot,\bar{q}/p}(x,\boldsymbol{k}^2_T)
   	\\&=-\Delta^N f_{\bar{q}/p}^{(1)}(x).
\end{align}
In of Fig.~\ref{f4}, the first $k_T$-moments of the $\bar{u}$ and $\bar{d}$ Sivers distribution functions are plotted by the solid line and the dashed line, respectively.
The left figure show the result from Eq.~(\ref{eq25}), while the right figure shows the result from Eq.~(\ref{eq33}).
Again, we find that the first $k_T$-moments for $\bar{u}$ and $\bar{d}$ in our model are both negative. 
The size of the moments is around $[0.003, 0.004]$ in the region $0.04<x<0.15$, where the magnitude has the largest value.
In the region $x>0.4$, the moments are negligible compared to those in the small $x$ region. 
Generally, the moment of $\bar{d}$ is larger than that of $\bar{u}$.
The reason of this phenomenon is that in our model the magnitude of $f_{1T}^{\perp,\pi^-/p}(x)$ is larger than that of $f_{1T}^{\perp,\pi^+/p}(x)$.
This is similar to the case of the unpolarized sea distribution, for which the distribution of $\bar{d}$ is larger than that of $\bar{u}$ in the intermediate $x$ region.
Both the phenomena can be explained in the meson-baryon fluctuation model, that is, the possibility of the fluctuation $p\to \pi^+ n$ is larger than the possibility of the fluctuation $p\to \pi^- \Delta^{++}$.

Finally, we would like to compare our model result with the model calculations or extractions from other groups~\cite{Bury:2021sue,he2019sivers,bacchetta2011constraining,anselmino2017study}.
In Ref.~\cite{he2019sivers}, the sea quark Sivers functions are calculated with the chiral model.
In that model, the Sivers function of $\bar{u}$ is negative and that of $\bar{d}$ is positive.
In addition, the size of the first $k_T$ moment timed with $x$ is 0.004 at most at the scale $Q=0.63$ GeV.
In Ref.~\cite{bacchetta2011constraining}, the Sivers functions including those of $\bar{u}$ and $\bar{d}$ are obtained by assuming a connection between generalized parton distribution $E$ and the Sivers function and by fitting the SSA data in SIDIS process.
In this study, the size of $xf_{1T}^{\perp (1)}$(x) for $\bar{u}$ and $\bar{d}$ is 0.006 at most, which is similar to the results in Ref.~\cite{he2019sivers}; However, their signs are opposite to those in Ref.~\cite{he2019sivers}, and the $x$-dependence are also different. 
A recent extraction~\cite{Bury:2021sue} which implements TMD evolution shows that the sign of the sea quark Sivers function is negative (In this extraction the $\bar{u}$ and $\bar{d}$ quarks are not distinguished).
This result is consistent with our model calculation. 
In general, the size of our results is similar with those in Refs.~\cite{he2019sivers,bacchetta2011constraining,anselmino2017study,Bury:2021sue}.
Hopefully, the reasons for these differences can be checked through theoretical and experimental analysis in the future.

\section{CONCLUSION}\label{Sec5}

In this work, we studied the sea quark ($\bar{u}$ and $\bar{d}$) Sivers function of the proton using a light-cone model.
In the model the proton can fluctuate to the Fock state formed by a pion meson and a baryon $B$, where the pion meson is composed by the $q\bar{q}$ pair.
We derive the light-cone wavefunction of this composite system in terms of $q \bar{q} B$.
Using the overlap representation of LCWFs, we calculated the unpolarized sea quark distribution function $f_{1}^{\bar{q}/p}(x,\boldsymbol{k}_T)$ and the sea quark Sivers function $f_{1T}^{\bot,\bar{q}/p}(x,\boldsymbol{k}^2_T)$.
In the calculation, we adopt the dipole form factors for both the nucleon-meson-baryon vertex and the meson-quark-antiquark vertex.
The values of the parameters in the model are fixed by fitting the model result $f_{1}^{\bar{q}/\pi}(x)$ and $f_{1}^{\bar{q}/p}(x)$ to GRV LO parametrization and MSTW2008 parametrization, respectively.
The calculation of the sea quark Sivers functions can be simplified by assuming the transverse momentum $\bm l_T$ does not go through the pion-quark-antiquark vertex, which means that the final-state interaction occurs between the pion meson as a whole and the spectator.
In this scenario the sea quark Sivers function can be expressed as the convolution of $f_{1}^{\bar{q}/\pi}$
and $f_{1}^{\bar{q}/\pi}$.
The numerical results show that the sign of the Sivers function of the $\bar{u}$ and $\bar{d}$ are negative in the entire $x$ and $k_T$ region.
For the first transverse moment of the sea quark Sivers function, the The size is 0.004 at most.
We also compared our model results with the recent extraction of $xf_{1T}^{\bot(1),\bar{q}/p}(x)$ and find some similarities and differences between them which need to be further classified by future theoretical studies and experimental measurement.
As an extension, our model can be used to estimate the other sea quark TMD distributions in leading-twist.
In conclusion, our study provide useful information of the sea quark Sivers function in a proton. Further study is needed in order to provide more stringent constraints on the sea quark Sivers functions.

\section*{Acknowledgements}
This work is partially supported by the National Natural Science Foundation of China under grant number 12150013.

\bibliography{REV}
\end{document}